# Modeling of asymmetric giant magnetoimpedance in amorphous ribbons with a surface crystalline layer

N.A. Buznikov [a,b,*], CheolGi Kim [b], Chong-Oh Kim [b], Seok-Soo Yoon [b,c]

[a] *Institute for Theoretical and Applied Electrodynamics, Russian Academy of Sciences, Moscow 125412, Russian Federation*

[b] *Research Center for Advanced Magnetic Materials, Chungnam National University, Daejeon 305-764, Republic of Korea*

[c] *Department of Physics, Andong National University, Andong 760-749, Republic of Korea*

**Abstract**

A model describing the asymmetric giant magnetoimpedance (GMI) in field-annealed amorphous ribbons is proposed. It is assumed that the ribbon consists of an inner amorphous core and surface hard magnetic crystalline layers. The model is based on a simultaneous solution of linearizied Maxwell equations and Landau–Lifshitz equation. The coupling between the surface layers and the amorphous core is described in terms of an effective bias field. Analytical expressions for the frequency and field dependences of the ribbon impedance are found. The calculated dependences are in a qualitative agreement with results of experimental studies of the high-frequency asymmetric giant GMI in field-annealed amorphous ribbons.

---

[*] Corresponding author. *E-mail address:* n_buznikov@mail.ru

# 1. Introduction

The giant magnetoimpedance (GMI) effect implies a significant change in the impedance of a soft magnetic conductor with the variation of an external magnetic field. The effect has been observed in different soft magnetic materials and has attracted a great deal of interest due to a possible use in various applications (see, e.g., Refs. [1–3] and references therein). The sensitivity and linearity for the magnetic field are the most important parameters in practical applications of GMI for magnetic sensors. An asymmetric behavior of the GMI profile is desirable because of this requirement. In this connection, much attention has been paid recently to the asymmetric GMI effect.

The asymmetric GMI has been observed first for twisted Co-based amorphous wires with DC bias current superimposed on the driving current [4]. A similar effect has been studied later in as-quenched amorphous ribbons [5] and wires [6]. Another method of producing the asymmetric GMI profile consists of applying an axial AC bias field to a sample [7]. However, these two methods have some limitations in applications, such as electrical power consumption.

Recently, a very large asymmetric GMI profile has been observed in Co-based amorphous ribbons annealed in air, in the presence of a weak magnetic field [8–12]. It has been shown that the asymmetry of the GMI profile is related to the hard magnetic crystalline layer, which appears due to the surface crystallization of a ribbon after the field annealing [9,13,14]. Although the effect has been studied experimentally quite well, the satisfactory theoretical explanation is still missing. An attempt to explain the asymmetric GMI in field-annealed ribbons in the framework of the quasi-static model [15] has not been very successful [16]. This model describes some basic features of the asymmetric GMI, however, it cannot explain the frequency dependence of the impedance. Moreover, the effect of the crystalline layer thickness is not taken into account in the model. More recently, a phenomenological model of the asymmetric GMI based on a solution of Maxwell equations and Landau–Lifshitz equation for semi-infinite ribbon with surface spin-pinning condition has been proposed [17]. Taking into account the exchange interactions and neglecting the anisotropy field, the problem of the impedance calculation reduces to a numerical solution of a differential equation for the magnetic field distribution in the ribbon. The model describes qualitatively the asymmetric GMI observed in experiments, however, a complex numerical procedure to calculate the impedance is required.

In this paper, we propose another model, which allows one to explain main features of the high-frequency asymmetric GMI in field-annealed ribbons. The model is based on a simultaneous solution of Maxwell equations together with Landau–Lifshitz equation for a ribbon of a finite thickness consisting of an amorphous core and two outer crystalline layers. The coupling between the surface layers and the amorphous core is considered in terms of an effective bias field. Neglecting a domain structure and exchange interactions, analytical expression for the impedance is obtained, and the field and frequency dependences of the impedance are analyzed.

## 2. Ribbon impedance

Let us consider a ribbon of thickness $D$ consisting of an inner amorphous core of thickness $d$ and two outer crystalline layers at the ribbon surface occurring after field annealing. The AC current $I=I_0\exp(-i\omega t)$ flows along the ribbon (along $z$-axis), and the external DC magnetic field $H_e$ is parallel to the current. The coordinate system is chosen so that the AC field induced by the current is parallel to the $y$-axis. A sketch of the coordinate system is shown in Fig. 1. It is assumed that the ribbon length and width are much higher than its thickness. For this reason, neglecting edge effects, we consider that the fields depend only on the coordinate perpendicular to the ribbon plane ($x$-coordinate).

The calculation of the impedance is based on a solution of Maxwell equations for the electric and magnetic fields together with Landau–Lifshitz equation of motion for the magnetization. An analytical treatment is possible in a linear approximation with respect to the time-variable parameters and under assumption of a local relationship between the magnetic field and the magnetization. In general, the domain-wall motion and the magnetization rotation process cause the permeability. Further, we neglect a domain structure of the ribbon and assume that the effective permeability is due to the magnetization rotation only. This assumption is known to be valid for not too low frequencies.

In the outer crystalline layers, $d/2<|x|<D/2$, the field annealing induces unidirectional anisotropy [8,9,12]. The value of the unidirectional anisotropy field in the crystalline layers $H_u$ is sufficiently high and of the order of several hundreds of Oe [12]. It is assumed further that the unidirectional anisotropy field in the outer layers has the constant angle $\varphi$ with respect to the transverse direction (see Fig. 1). Note that the direction of $H_u$ differs from that of the annealing field, which may be attributed to the influence of the amorphous core on the crystallization process in the surface layers.

The distribution of the easy axes in the amorphous core, $|x|<d/2$, depends on the stresses induced in the ribbon during the fabrication process and may vary significantly over the ribbon volume. We assume for simplicity that the amorphous core has the uniaxial anisotropy, and the anisotropy axis makes the angle $\psi$ with the transverse direction in the whole region $|x|<d/2$. The magnetostatic coupling between amorphous and crystalline phases induces the effective bias field $H_b$ in the amorphous region, with the bias field being in the opposite direction to the unidirectional anisotropy field $H_u$ [12] (the angle of the bias field with respect to the transverse direction is $\varphi+\pi$, see Fig. 1). To obtain analytical results, we assume that the effective bias field does not vary over the amorphous region.

To calculate the permeability we neglect the demagnetizing fields and the contribution of the exchange energy. The permeability tensor $\hat{\mu}$ for both the inner amorphous region and outer crystalline layers can be found by means of the well-known procedure of the solution of linearized Landau–Lifshitz equation. In general, the permeability is represented by a non-diagonal tensor [18–20]. For further analysis, we use the values of the effective transverse permeability in the amorphous core and outer crystalline layers. In the amorphous region, the effective transverse permeability $\mu_1$ is given by [20]

$$\mu_1 = 1 + \frac{\omega_m(\omega_m + \omega_1 - i\alpha\omega)\sin^2\theta}{(\omega_m + \omega_1 - i\alpha\omega)(\omega_2 - i\alpha\omega) - \omega^2}, \tag{1}$$

$$\omega_m = \gamma 4\pi M,$$
$$\omega_1 = \gamma[H_a\cos^2(\theta-\psi) - H_b\cos(\theta-\varphi) + H_e\sin\theta], \tag{2}$$
$$\omega_2 = \gamma[H_a\cos\{2(\theta-\psi)\} - H_b\cos(\theta-\varphi) + H_e\sin\theta].$$

Here $M$ is the saturation magnetization, $H_a$ is the uniaxial anisotropy field, $\gamma$ is the gyromagnetic constant, $\alpha$ is the Gilbert damping parameter and $\theta$ is the equilibrium angle between the magnetization vector and the transverse direction. The angle $\theta$ can be found by minimizing the free energy, which can be presented as a sum of the uniaxial anisotropy energy, bias field energy and Zeeman energy. The minimization procedure results in the equation for the equilibrium angle

$$H_a \sin(\theta - \psi)\cos(\theta - \psi) - H_b \sin(\theta - \varphi) - H_e \cos\theta = 0. \tag{3}$$

Since the unidirectional anisotropy field $H_u$ is high, the transverse permeability $\mu_2$ in the outer layers is almost independent of the external field at low $H_e$ and can be expressed as

$$\mu_2 = 1 + \frac{\omega_m(\omega_m + \omega_3 - i\alpha\omega)\sin^2\varphi}{(\omega_m + \omega_3 - i\alpha\omega)(\omega_3 - i\alpha\omega) - \omega^2},$$
$$\omega_3 = \gamma H_u. \tag{4}$$

To simplify calculations, we assume that the saturation magnetization in the crystalline layers is equal to that in the amorphous core. Note that the modifications taking into account the different values of the saturation magnetization in the amorphous and crystalline phases can be readily made in the framework of the model.

The distribution of the electric and magnetic fields in the ribbon satisfies Maxwell equations:

$$\text{curl}\,\mathbf{e} = (i\omega/c)\hat{\mu}(x)\mathbf{h},$$
$$\text{curl}\,\mathbf{h} = 4\pi\sigma(x)\mathbf{e}/c, \tag{5}$$

where $\mathbf{e}$ and $\mathbf{h}$ are the AC electric and magnetic fields, respectively, $c$ is the velocity of light and $\sigma(x)$ is the conductivity. It is assumed that the conductivity may differ for the inner amorphous core and the outer crystalline layers: $\sigma(x)=\sigma_1$ at $|x|<d/2$, and $\sigma(x)=\sigma_2$ at $d/2<|x|<D/2$.

Since we assume that the fields depend only on the coordinate perpendicular to the ribbon plane, Maxwell equations reduce to two coupled differential equations for the components of the AC magnetic field $h_y$ and $h_z$ [17,19,20]. In this paper, we neglect for simplicity the longitudinal AC field $h_z$. This assumption can be made, since $h_y \gg h_z$ in ribbons excited by the AC current flowing along $z$-axis. Then, the problem is simplified, and the solution of Eqs. (5) for the inner amorphous core, $|x|<d/2$, has the form

$$e_z^{(1)} = A\cosh(\lambda_1 x),$$
$$h_y^{(1)} = (4\pi\sigma_1/c\lambda_1)A\sinh(\lambda_1 x), \quad (6)$$
$$\lambda_1 = (1-i)/\delta_1, \quad \delta_1 = c/(2\pi\sigma_1\omega\mu_1)^{1/2},$$

where $A$ is a constant.

In the regions of the crystalline layers, $d/2<|x|<D/2$, the electric and magnetic fields are given by

$$e_z^{(2)} = B\cosh(\lambda_2 x) + C\sinh(\lambda_2 x),$$
$$h_y^{(2)} = (4\pi\sigma_2/c\lambda_2)[B\sinh(\lambda_2 x) + C\cosh(\lambda_2 x)], \quad (7)$$
$$\lambda_2 = (1-i)/\delta_2, \quad \delta_2 = c/(2\pi\sigma_2\omega\mu_2)^{1/2}.$$

The constants $B$ and $C$ in Eqs. (7) can be determined from the corresponding boundary conditions at the surface of the inner amorphous layer:

$$e_z^{(1)}(d/2) = e_z^{(2)}(d/2),$$
$$h_y^{(1)}(d/2) = h_y^{(2)}(d/2). \quad (8)$$

Taking into account Eqs. (6) and (7), we find for these values

$$B/A = \cosh(\lambda_1 d/2)\cosh(\lambda_2 d/2) - (\sigma_1\lambda_2/\sigma_2\lambda_1)\sinh(\lambda_1 d/2)\sinh(\lambda_2 d/2),$$
$$C/A = (\sigma_1\lambda_2/\sigma_2\lambda_1)\sinh(\lambda_1 d/2)\cosh(\lambda_2 d/2) - \cosh(\lambda_1 d/2)\sinh(\lambda_2 d/2). \quad (9)$$

The impedance $Z$ of the ribbon can be obtained by means of the relation

$$Z = le_z^{(2)}(D/2)/I = (2\pi l/cw)e_z^{(2)}(D/2)/h_y^{(2)}(D/2), \quad (10)$$

where $l$ and $w$ are the ribbon length and width, respectively. Using Eqs. (7), (9) and (10), we find for the impedance

$$\frac{Z}{R_{DC}} = (\lambda_1\sigma_2/2\sigma_1)[D - d(1-\sigma_1/\sigma_2)]$$
$$\times \frac{\coth(\lambda_1 d/2) + (\sigma_1\lambda_2/\sigma_2\lambda_1)\tanh(\lambda_2\{D-d\}/2)}{1 + (\sigma_2\lambda_1/\sigma_1\lambda_2)\coth(\lambda_1 d/2)\tanh(\lambda_2\{D-d\}/2)}, \quad (11)$$

where $R_{DC}=(l/w)[\sigma_2 D+(\sigma_1-\sigma_2)d]^{-1}$ is the DC ribbon resistance. Therefore, the dependence of the ribbon impedance both on the external magnetic field and the current frequency can be calculated by means of Eqs. (1)–(4), (6), (7) and (11). Note that at $d=D$ and $\sigma_1=\sigma_2$, Eq. (11)

coincides with the usual expression for the impedance of amorphous ribbon without crystalline layers [3,18,21].

## 3. Results and discussion

Fig. 2 shows the ribbon impedance as a function of the external magnetic field $H_e$ at the current frequency $f = \omega/2\pi = 5$ MHz for different values of the bias field $H_b$. The saturation magnetization, the unidirectional anisotropy field, the uniaxial anisotropy field and the damping parameter of the ribbon are taken as $4\pi M = 7000$ Gs, $H_u = 300$ Oe, $H_a = 2$ Oe and $\alpha = 0.1$, respectively. For simplicity we assume that the conductivity of the crystalline layers is equal to that of the amorphous core, $\sigma_1 = \sigma_2$.

It follows from Fig. 2 that impedance profiles show the asymmetric two-peak behavior. The asymmetry growths with the bias field $H_b$, the negative field peak decreases and the positive field peak increases. The peak values of $H_e$ shift to the direction of the annealing field with the increase of the effective bias field. Fig. 2 illustrates also the effect of the unidirectional anisotropy field angle $\varphi$ on the impedance profile. It is seen from Fig. 2 that the asymmetry of the impedance profile increases with the deviation of the unidirectional anisotropy field from the annealing field direction. Note that it is assumed that the direction of the anisotropy field $H_u$ in the surface layers may differ from that of the annealing field. This fact is attributed to the influence of the uniaxial anisotropy in the amorphous phase on the surface crystallization process. As a result, the unidirectional anisotropy field $H_u$ deviates from the ribbon axis, and the angle $\varphi$ lies within the range of the angles of the uniaxial anisotropy field in the amorphous core and the annealing field.

Shown in Fig. 3 is the field dependence of the ribbon impedance at fixed bias field for different frequencies and two values of the crystalline layer thickness. It follows from Fig. 3 that the relative difference between peaks decreases with the increase of the current frequency. At the same time, the impedance increases with the frequency, which is related to the decrease of the skin depth $\delta_1$ in the amorphous layer. It is seen also from Fig. 3 that the impedance decreases with the increase of the outer layer thickness, and the field dependence of the impedance remains the same. This fact has a clear physical meaning. With the growth of the thickness of the crystalline layer, its relative contribution to the impedance increases. However, since the permeability of the hard outer layers is independent of the external field, the crystalline layers do not affect the field dependence of the impedance. Note that the calculated field and frequency dependences of the impedance are in a qualitative agreement with high-frequency GMI profiles observed in the experiments [8–12].

In conclusion of this section, it should be noted that in the model proposed, the permeability is determined only by the magnetization rotation process. At sufficiently low frequencies about 1 MHz and less, the contribution from the domain-walls motion to the transverse permeability should be taken into account [2,3,18]. The domain-walls motion may result in a drastic step-like change in the impedance at low frequencies (so-called "GMI valve") [12,22]. The analysis of the influence of the domain structure on the GMI response will be published elsewhere. Moreover, to calculate the effective transverse permeability we neglect the coordinate dependence of the effective bias field, which allows one to obtain analytical expressions for the ribbon impedance. Taking into account the spatial distribution of the bias field in the amorphous layer may be essential for a detail description of the asymmetric GMI profile. Note also that we assume in calculations that the saturation magnetization, conductivity and damping parameter do not change after the field annealing. However, corresponding modifications can be readily made in the framework of the present approach. Nevertheless, even a simplified model explains quite well the observed high-frequency asymmetric GMI in amorphous ribbons.

## 4. Conclusions

In this paper, the rotational model to calculate the impedance of the ribbon consisting of an amorphous core and two hard magnetic crystalline layers is developed. In the framework of the model, analytical expressions for the impedance are found taking into account the influence of the effective bias field occurring due to the interaction between the amorphous and the crystalline layers. The results of calculations allow one to describe main features of the field and frequency dependences of the asymmetric GMI profile observed in field-annealed amorphous ribbons at high frequencies [8−12]. The results obtained may be useful to develop materials with asymmetric GMI profile for sensitive field sensors.

**Acknowledgements**

This work was supported by the Korea Science and Engineering Foundation through ReCAMM. N.A. Buznikov would like to acknowledge the support of the Brain Pool Program.

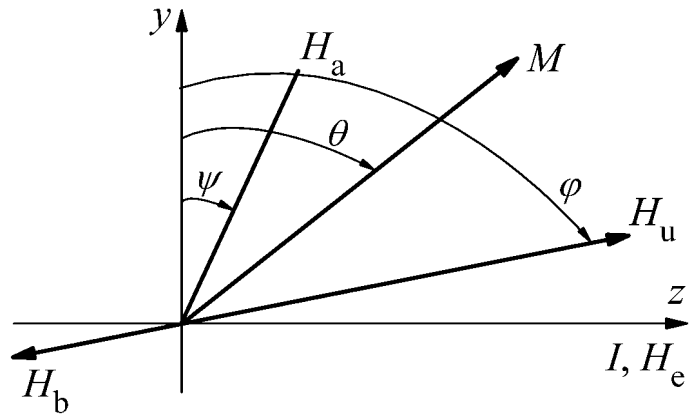

Fig. 1. A sketch of the coordinate system used for analysis. All the vectors lie within $y$–$z$ plane. The bias field $H_b$ in the amorphous region is in the opposite direction to the unidirectional anisotropy field $H_u$ in the crystalline layer.

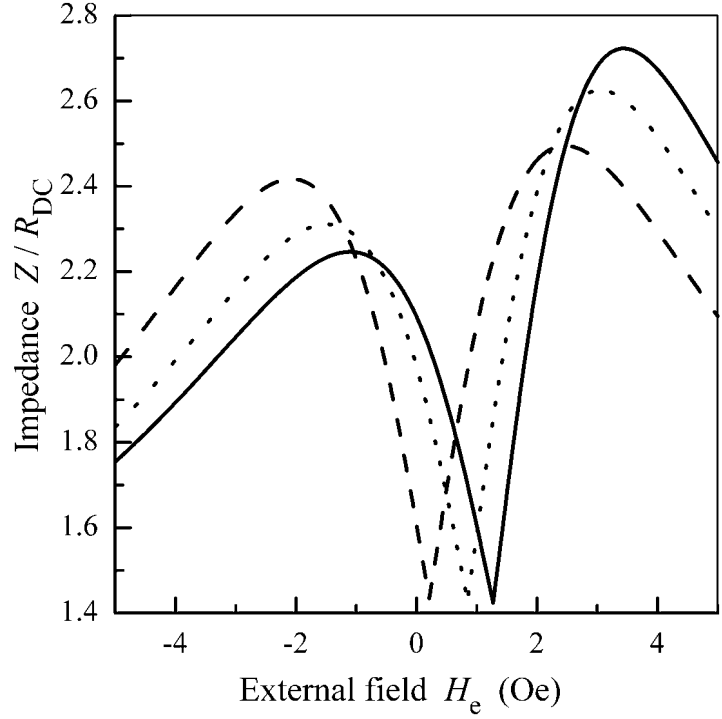

(a)

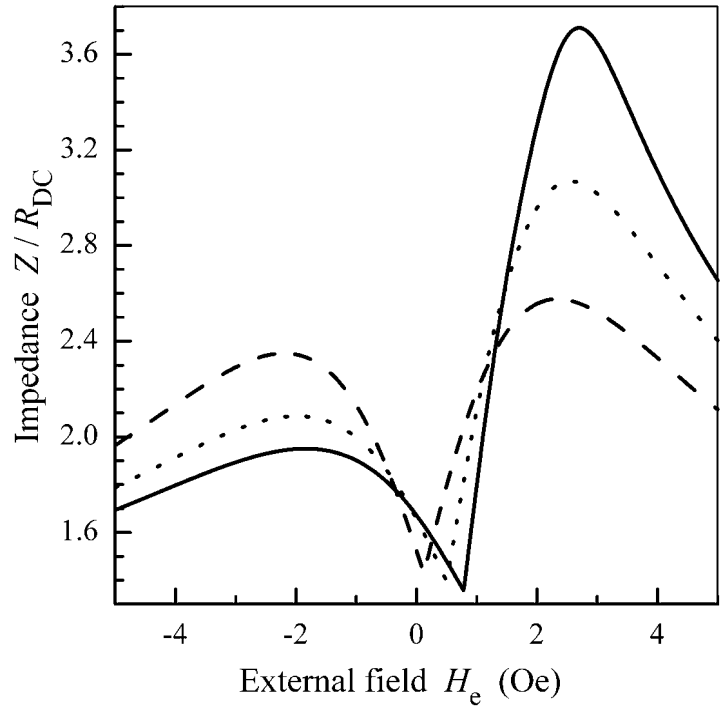

(b)

Fig. 2. Dependences of $Z$ on $H_e$ at $f=5$ MHz and different $\varphi$: (a) $\varphi=0.45\pi$; (b) $\varphi=0.35\pi$. Dashed lines, $H_b=0.25$ Oe; dotted lines, $H_b=1$ Oe; solid lines, $H_b=1.5$ Oe. Parameters of ribbon used for calculations are $D=20$ μm, $d=18$ μm, $\sigma_1=\sigma_2=10^{16}$ s$^{-1}$, $4\pi M=7000$ Gs, $H_a=2$ Oe, $H_u=300$ Oe, $\alpha=0.1$, $\psi=0.15\pi$.

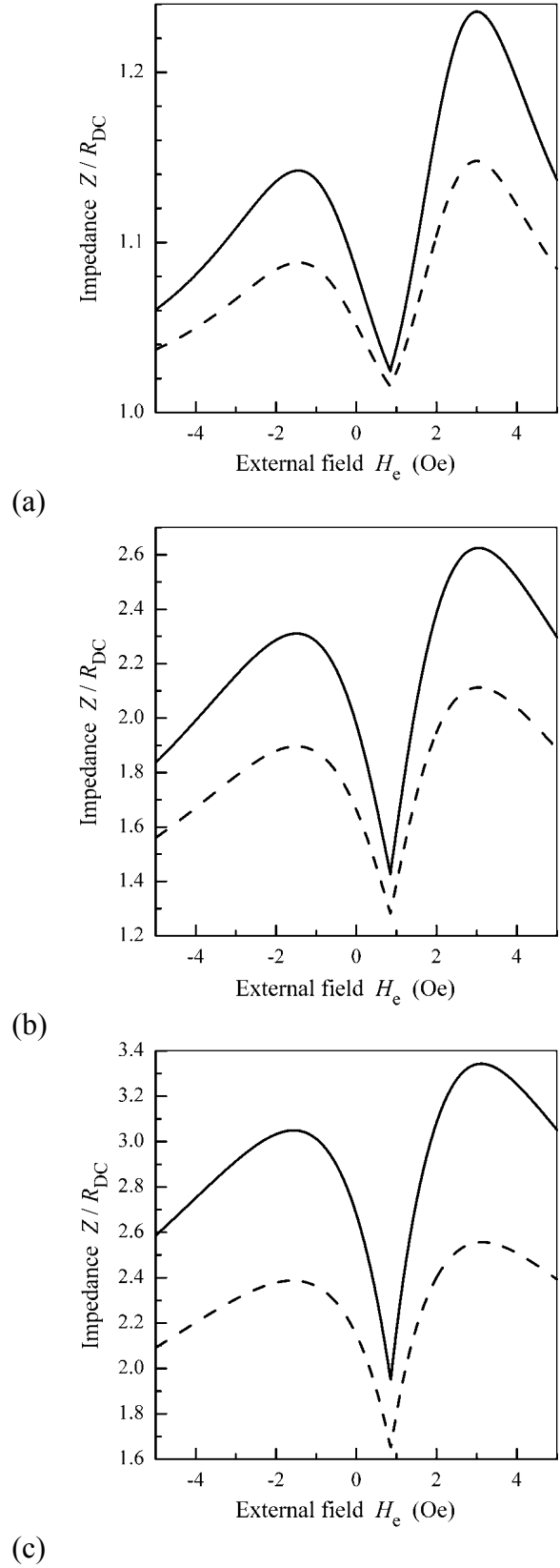

Fig. 3. Dependences of $Z$ on $H_e$ at $H_b=1$ Oe and different $f$: (a) $f=1$ MHz; (b) $f=5$ MHz; (c) $f=10$ MHz. Solid lines, $d=18$ μm; dashed lines, $d=16$ μm. Parameters of ribbon used for calculations are $D=20$ μm, $\sigma_1=\sigma_2=10^{16}$ s$^{-1}$, $4\pi M=7000$ Gs, $H_a=2$ Oe, $H_u=300$ Oe, $\alpha=0.1$, $\psi=0.15\pi$, $\varphi=0.45\pi$.